# Evaluation of state-of-the-art deep learning models in the segmentation of the heart ventricles in parasternal short-axis echocardiograms

**Julian R. Cuellar,**[a,*] **Vu Dinh,**[b] **Manjula Burri,**[c] **Julie Roelandts,**[d] **James Wendling,**[d] **Jon D. Klingensmith,**[a]

[a] Southern Illinois University Edwardsville, 61 Circle Dr, Edwardsville, IL, USA 62025
[b] Russel H. Morgan Department of Radiology and Radiological Science, The Johns Hopkins University School of Medicine, 1800 Orleans St, Baltimore, MD, USA 21287
[c] Columbus Regional Hospital, 2400 17th St, Columbus, IN, USA 47201
[d] Department of Diagnostic Medical Sonography, St. Louis Community College, St. Louis, Missouri, USA 63044

**Abstract**

**Purpose:** Previous studies on echocardiogram segmentation are focused on the left ventricle in parasternal long-axis views. In this study, deep-learning models were evaluated on the segmentation of the ventricles in parasternal short-axis echocardiograms (PSAX-echo). Segmentation of the ventricles in complementary echocardiogram views will allow the computation of important metrics with the potential to aid in diagnosing cardio-pulmonary diseases and other cardiomyopathies. Evaluating state-of-the-art models with small datasets can reveal if they improve performance on limited data.

**Approach:** PSAX-echo were performed on 33 volunteer women. An experienced cardiologist identified end-diastole and end-systole frames from 387 scans, and expert observers manually traced the contours of the cardiac structures. Traced frames were pre-processed and used to create labels to train 2 specific-domain (Unet-Resnet101 and Unet-ResNet50), and 4 general-domain (3 Segment Anything (SAM) variants, and the Detectron2) deep-learning models. The performance of the models was evaluated using the Dice similarity coefficient (DSC), Hausdorff distance (HD), and difference in cross-sectional area (DCSA).

**Results:** The Unet-Resnet101 model provided superior performance in the segmentation of the ventricles with 0.83, 4.93 pixels, and 106 pixel$^2$ on average for DSC, HD, and DCSA respectively. A fine-tuned MedSAM model provided a performance of 0.82, 6.66 pixels, and 1252 pixel$^2$, while the Detectron2 model provided 0.78, 2.12 pixels, and 116 pixel$^2$ for the same metrics respectively.

**Conclusions:** Deep-learning models are suitable for the segmentation of the left and right ventricles in PSAX-echo. This study demonstrated that specific-domain trained models such as Unet-ResNet provide higher accuracy for echo segmentation than general-domain segmentation models when working with small and locally acquired datasets.

*Julian R. Cuellar, E-mail: jrenecb@gmail.com, jucuell@siue.edu

## 1 Introduction

Echocardiography (echo) is a widely used imaging modality for the assessment of the anatomical structure and function of the heart. Compared with other imaging modalities like magnetic resonance imaging (MRI) and computed tomography (CT), echo has several advantages - it is more affordable, portable, noninvasive, and provides real-time acquisition of two-dimensional images of the patient's heart.[1,2] The initial step in quantitative assessment using echo typically involves the segmentation of the left ventricle (LV) and right ventricle (RV). Measuring the LV and RV facilitates the extraction of clinical parameters such as area, volume, and mass of the LV and RV, ejection fraction (EF), and myocardial mass.[3] In clinical practice, segmentation of the ventricles is performed using manual delineation at end-diastole (ED) and end-systole (ES) by a cardiologist or expert user. However, manual segmentation is a tedious and laborious process and suffers from inter- and intra- observer variability.[4] Therefore, an automated process for the segmentation of the two ventricles is preferred and can obviate the limitations of the traditional manual process. However, the accuracy of these segmentation tasks often suffers from several intrinsic drawbacks of ultrasound imaging such as the low contrast between the structures of interest due to low signal-to-noise ratio, presence of speckle noise, shadows, signal and edge dropout, and motion artifacts caused by patient or probe movement.

Many semi-automated or fully-automated approaches have been proposed for the segmentation of the LV and RV using either traditional machine-learning methods or deep neural networks. Traditional machine-learning-based segmentation methods have been constrained to the segmentation of single anatomic structures and are cumbersome for multiple objects, especially when segmenting them concurrently.[5–7] On the other hand, multi-tissue segmentation can be feasible using deep-learning (DL) approaches. Methods based on deep neural networks started proliferating after a well-known convolutional neural network (CNN) architecture called a U-Net was introduced for biomedical image segmentation in 2015.[8] In recent studies, many researchers have experimented with U-Net-derived approaches[9] and used the CAMUS dataset[10] as training input. Among these studies, the MFP-Unet and Unet-ResNet34 proposed by Moradi et al. and Zyuzin et al., respectively, were trained on the CAMUS dataset and achieved high accuracy in segmenting the LV with Dice similarity coefficient (DSC) scores of 0.95 and Hausdorff distances (HD) of 3.49.[11,12] The MV-RAN model proposed by Ming Li et al. processes echocardiographic sequences and analyses the full cardiac cycle. This model was trained using a private dataset of A2C, A3C, A4C, and the CAMUS dataset and achieved a DSC score of 0.92 and an HD of 6.06.[13] A model called GL-Fusion that explores the cyclic relations on echocardiograms was presented by Ziyang Zheng et al. The model was trained with a collection of LV multi-view echo images called MvEVD and achieved a DSC score of 0.83.[14] Studies using other public datasets such as EchoNet-Pediatric and EchoNet-Dynamic only report results of the segmentation of LV with DSC scores of 0.94 and 0.90 respectively using the EchoSegDiff model.[15] The LV has been a target structure in these studies due to its important correlation to heart failure and is often segmented with high accuracy.

In contrast to approaches for segmentation of the LV, fewer attempts have been made to segment the RV in PSAX-echo. This is despite the importance of RV measurements in detecting cardiopulmonary diseases, coronary heart diseases, and cardiomyopathy.[16–18] Various factors limit the number of studies for segmentation of the RV in ultrasound images of the heart, including irregular and inconsistent geometry, difficulty in differentiating the chamber and the myocardium of the RV due to similar grayscale brightness, and generally poorer image quality in echocardiographic images.[19,20] For other modalities, several researchers have investigated the potential of deep neural networks based on the U-Net architectures for the segmentation of the RV in MRI and CT images.[21–23] In studies using PSAX-echo images, the segmentation of the LV is presented with DSC scores of 0.83, 0.75, 0.92, and 0.88.[24–27] Other studies report the segmentation of cardiac structures including the aortic valve and the coronary artery.[28,29] The Unet-ResNet34 model was also used to segment the LV and RV in PSAX-echo in one of our previous studies with a small data set of 506 expert-traced images. It showed promise for this segmentation task providing average DSC scores of 0.94, 0.89, and 0.88, and Hausdorff distances (HD) of 4.81, 4.71, and 7.89 for the three boundaries, respectively.[2] This study investigated the potential to segment both ventricles simultaneously. We improved the Hausdorff distances while achieving DSC scores comparable to the models presented in the review. A larger dataset of ~1700 images was used and the performance of UNET-ResNet architectures was tested against novel models.

Deep-learning models can facilitate the development of fully-automatic algorithms for the segmentation of structures of interest. These models are highly application-dependent and modifications of the architectural design of the models are necessary to meet the specific requirements of the application. During the last few years, well-known research groups such as Microsoft Research (MR), and the Facebook AI Research (FAIR) team at Meta, have developed robust state-of-the-art (SOTA) AI models trained in large-scale datasets such as ImageNet, COCO, SA-1B, Pascal VOC, ADE20K, and Cityscapes. The high performance and capabilities of models based on ResNets from MR,[30–32] the Segment Anything Model (SAM)[33] and the Detectron2[34] libraries from FAIR have captured the attention of researchers in medical imaging segmentation. Using AI training techniques such as transfer learning and fine-tuning, Jun Ma et al. developed MedSAM,[35] a foundation model for medical imaging segmentation tasks based on SAM. Table 1 presents the main characteristics of these DL models and the description of the datasets used for their training.

**Table 1** Characteristics of SOTA image segmentation models.

| Model | Architecture | Training Data | Dataset Size | Use Cases |
|---|---|---|---|---|
| Unet-ResNet | Encoder-Decoder | ImageNet | ~ 1.2 million images<br>~ 1000 categories | Biomedical image segmentation |
| Detectron2 | Modular, Region-based | COCO<br><br>Pascal VOC | ~ 123,000 images<br>~ 900,000 annotations<br>~ 10,000 images<br>~ 25,000 annotations | Object detection, instance segmentation, panoptic segmentation |
| SAM | Complex, Foundation Model | COCO<br>ADE20K<br><br>Open Images<br><br>SA-1B | --<br>~ 28,000 images<br>~ 700,000 annotations<br>~ 1.0 million images<br>~ 2.7 million annotations<br>~ 11.0 million images<br>~ 1.0 billion annotations | General-purpose segmentation |
| MedSAM | Complex, Foundation Model | Finetuned SAM on large-scale medical image dataset | ~ 1.5 million images<br>~ 1.5 million annotations | Medical image segmentation |

PSAX-echo provides a cross-sectional view of the heart. This imaging technique is crucial for assessing various cardiac structures and functions, including left and right ventricular dimensions, wall thickness, and contractility. It is also used for examining the structure and function of the aortic and mitral valves, identifying congenital heart defects, including ventricular septal defects (VSDs), evaluating cardiac masses such as tumors or blood clots within the heart chambers, and ventricular dilation and wall motion abnormalities.[36] PSAX-echo also provides a view of the interventricular sulci, where the coronary arteries are located.[37] Coronary arteries in this area are surrounded by cardiac adipose tissue (CAT), which has been correlated with cardiovascular disease (CAD).[38–40] However, gray-scale echo images acquired by standard echocardiography,[41] and available in CAMUS,[42] EchoNet-Dynamic,[43] or other public datasets do not provide the raw radiofrequency (RF) data. By obtaining the RF data from echocardiograms, it becomes possible to apply advanced spectral analysis, potentially enabling cardiac tissue classification.[2,44] Segmentation of both ventricles in PSAX-echo is the first step toward identifying targeting areas for CAT. Processing the RF data in these regions will allow the identification and quantification of CAT, providing a valuable tool for assessing CAD risk using the real-time and cost-effective echocardiogram imaging modality.

In this study, our dataset of 387 PSAX-echo scans containing the raw RF ultrasound data was used to train the Unet-Resnet50 and Unet-Resnet101 models. These architectures became our specific-domain DL models. In addition, the Detectron2 and MedSAM general-domain models were fine-tuned with our dataset and their performance was evaluated for segmentation of the LV and RV. The fine-tuning process takes a pre-trained model and adjusts it for a specific task by continuing the training in a small dataset.[45] The performance of the original SAM and MedSAM models (without fine-tuning) was also tested in this study for a complete comparison of SOTA models in this challenging segmentation task.

## 2 Materials and Methods

### 2.1 Image Dataset

A total of 387 PSAX images of the heart were collected from 33 female volunteers aged 18 to 40 with a Body Mass Index (BMI) in the range of 30 to 39.99 kg/m$^2$. PSAX images from the base to the apex of the heart were acquired following 2 protocols. In protocol 1, PSAX images were taken at 3 view levels using the nominal angle of the US transducer. In protocol 2, the images were taken at 5 view levels and 3 transducer angles at each view level as shown in Table 2. Image and raw data were acquired using a Mindray Zonare ZS3 ultrasound system with a P4-1c phased-array transducer operating in harmonic mode at 3.5 MHz. In addition, the raw RF data from the transducer was base-band demodulated and decimated into in-phase and quadrature (IQ) data in complex format.[46] The IQ data consists of 80 temporal frames. Each frame has 201 A-lines and each A-line has 596 samples.

Table 2 Parasternal Short Axis image acquisition details.

| Protocol | View level | Transducer position(s) | Number of images | Number of Subjects |
|---|---|---|---|---|
| 1 | Mitral valve | Nominal | 1 | 9 |
|  | Papillary muscles |  | 1 |  |
|  | Near apex |  | 1 |  |
| 2 | Mitral valve | Nominal, rotate medial, rotate lateral | 3 | 24 |
|  | Chordae |  | 3 |  |
|  | Papillary muscles |  | 3 |  |
|  | Near apex |  | 3 |  |
|  | Near apex |  | 3 |  |

### 2.2 Gold Standard for Segmentation

A custom application was developed to use the raw RF data to reconstruct the traditional PSAX B-mode images to be used for annotation. The RF-to-B-mode conversion algorithm mimicked the Zonare system, but also included an additional Wavelet-based denoising algorithm to improve the quality of the images.[47] The application allowed one clinical cardiologist board-certified in echocardiography to identify end-diastolic (ED) and end-systolic (ES) frames among an 80-frame loop, and with two registered diagnostic cardiac sonographers, perform the manual tracing of the left endocardial border and left epicardial border of LV, and the right epicardial border of RV on the ED and ES images. The gold standard for the image segmentation in this study consisted of the ED/ES frame identified on each 80-frame loop, and the three experts' tracings of the cardiac structures. Our PSAX-echo dataset included 2322 labeled image pairs and the gold standard development is depicted in Fig. 1.

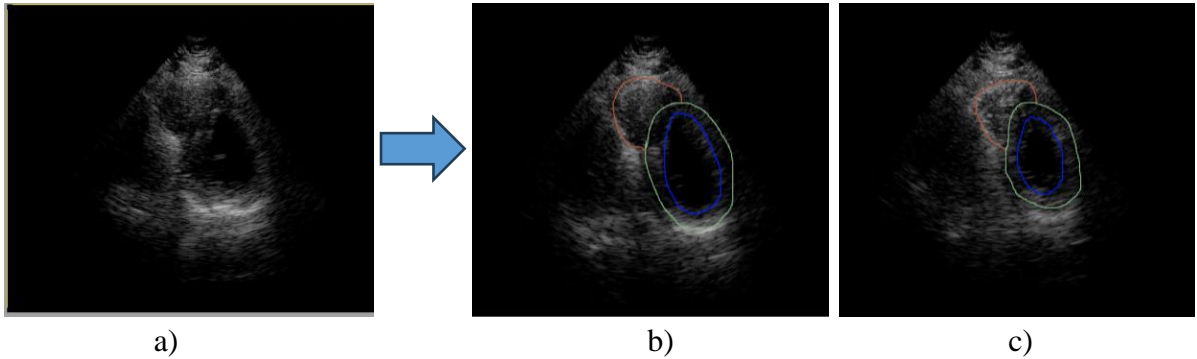

a)   b)   c)

**Fig. 1** Gold standard development. a) Echocardiogram with 80 temporal frames, b) Identification of End-Diastole frame, c) Identification of End-Systole frame. Experts traced the left myocardium, left epicardium, and right epicardium using blue, green, and red colors respectively.

## 2.3 Deep Learning Models for Image Segmentation

The segmentation of echo images was performed using SOTA DL models. These models include specific-domain deep U-Net architectures based on ResNet backbones (Unet-ResNet50 and Unet-ResNet101), as well as general-domain models such as the original SAM model developed by the FAIR team at Meta. Two SAM variations were also considered: MedSAM, a SAM model fine-tuned on a large-scale medical image dataset, and USmedSAM, a MedSAM model fine-tuned on our PSAX-echo image dataset. Additionally, the Detectron2 model was fine-tuned on our dataset.

### 2.3.1 Unet-Resnet models

The U-Net is a convolutional neural network (CNN) that was introduced in 2015.[8] The general architecture of the U-Net consists of encoding and decoding pathways with a depth of 5 as shown in Fig. 2 (left). The Unet-ResNet50 and Unet-ResNet101 have a Unet-based architecture, but they integrate residual blocks between the network layers in the encoding path.[48]

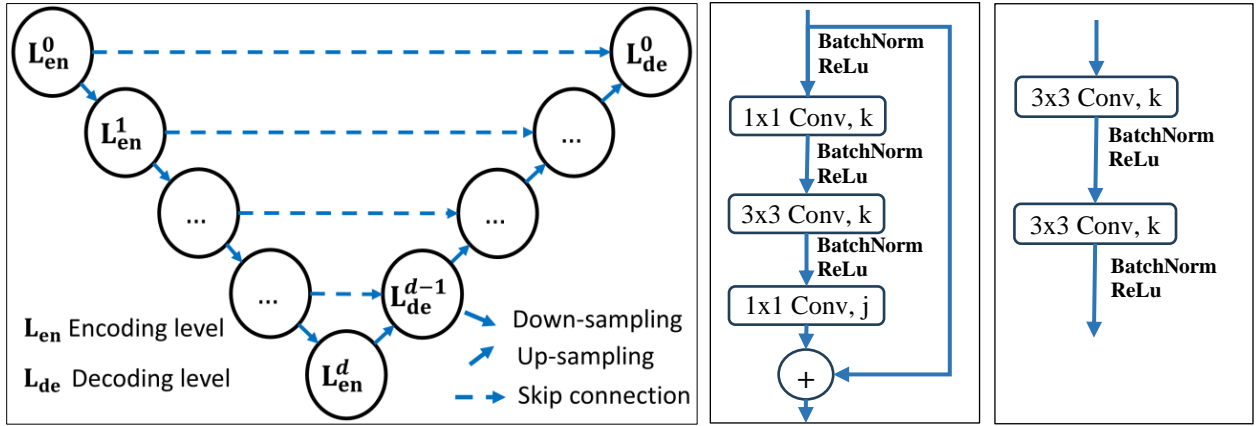

**Fig. 2** U-Net architecture (left), ResNet residual block (center), and decoder block (right). BatchNorm = Batch normalization operation, ReLU = Rectified linear unit, Conv = Convolutional layers.

A residual block is illustrated in Fig. 2 (center), it contains two 1x1, and one 3x3 convolutional layers with *k* and *j* output filters, three batch normalization layers, and three ReLU functions. After the third convolutional layer, the input of the residual block is added to its output. In the decoder path, decoder blocks with two 3x3 convolutional layers with *k* output filters, two batch normalization layers, and two ReLU functions as presented in Fig. 2 (right) are used. Table 3 provides a complete description of the ResNets network architecture.

The input image data has a size of 256x256x3. The first encoding level that operates on the input data uses two 5x5 convolutional layers and two Rectified Linear Unit (ReLU) activation functions. The input data is transformed into a feature map after this level. A 2x2 max pooling layer is applied to the output block at each step of the encoding path to retain important features of the input feature map, reduce the resolution of the feature map by half, and double the number of feature channels (see Table 3). At each decoding level, the feature map is first upscaled by a factor of 2 and concatenated to the output of the encoding block at the same level. The concatenated feature map is passed on to the next decoding block. After the 5 decoding levels, the feature map is used as input to a 1x1 convolutional layer with a SoftMax classifier to produce a probability map where each location contains a vector of probabilities representing each of the classes predicted by the model.

**Table 3** Layer details of the Unet-ResNet50, and the Unet-ResNet101 architectures. Details of ResNet residual and decoder blocks are depicted in Fig. 2 (center) and Fig. 2 (right). $L_{en}$ = encoding level, $L_{de}$ = decoding level, BatchNorm = batch normalization operation, ReLU = rectified linear unit, Conv = convolutional layers.

| Layer | Output size | ResNet50 | ResNet101 |
|---|---|---|---|
| $L_{en}^0$ | 256x256 | input | input |
| $L_{en}^1$ | 128x128 | Conv 5x5, /2<br>BatchNorm<br>ReLU | Conv 5x5, /2<br>BatchNorm<br>ReLU |
| $L_{en}^2$ | 64x64 | [residual block, k=64, j=256] x 3 | [residual block, k=64, j=256] x 3 |
| $L_{en}^3$ | 32x32 | [residual block, k=128, j=512] x 4 | [residual block, k=128, j=512] x 4 |
| $L_{en}^4$ | 16x16 | [residual block, k=256, j=1024] x 6 | [residual block, k=256, j=1024] x 18 |
| $L_{en}^5$ | 8x8 | [residual block, k=512, j=2048] x 3 | [residual block, k=512, j=2048] x 8 |
| $L_{de}^4$ | 16x16 | decoder block, k = 256 | decoder block, k = 256 |
| $L_{de}^3$ | 32x32 | decoder block, k = 128 | decoder block, k = 128 |
| $L_{de}^2$ | 64x64 | decoder block, k = 64 | decoder block, k = 64 |
| $L_{de}^1$ | 128x128 | decoder block, k = 32 | decoder block, k = 32 |
| $L_{de}^0$ | 256x256 | decoder block, k = 16 | decoder block, k = 16 |
| **Segmentation** | 256x256 | Conv 1x1, 4, SoftMax | |
| **Down-sampling between encoder block** | | Max Pooling 3x3, /2 after $L_{en}^1$ first conv layers in $L_{en}^3$-$L_{en}^5$ have stride 2 | Max Pooling 3x3, /2 after $L_{en}^1$ first conv layers in $L_{en}^3$-$L_{en}^5$ have stride 2 |
| **Up-sampling between decoder blocks** | | UpSampling repeats 2x2 | UpSampling repeats 2x2 |
| **Depth** | | 50 | 101 |

*2.3.2 Detectron2 model*

Detectron2 is a modular object detection library based on PyTorch. It is flexible and allows the high-quality implementation of SOTA object detection algorithms such as DensePose, panoptic feature pyramid networks, and variants of the Mask R-CNN.[34] In this study, from the Detectron2 library, a Mask R-CNN model with a ResNet50 backbone, and a feature pyramid network (FPN) for feature extraction was selected. This model was pre-trained by FAIR with the COCO dataset and we finetuned it on our PSAX-echo dataset.

The Mask R-CNN (Mask Region-based CNN) is a SOTA DL model for object detection and instance segmentation. It extends the Faster R-CNN architecture by adding a fully convolutional network (FCN) head that predicts a pixel-level segmentation mask for each detected object. This allows Mask R-CNN to not only localize objects within an image but also accurately delineate their boundaries. Two key elements of the Mask R-CNN architecture are the network head and the region proposal network (RPN). The RPN generates region proposals or bounding boxes that potentially contain objects. The network head consists of the branch that predicts the class for the proposed region (classification branch), and the branches that predict a pixel-level segmentation mask for each proposed region (mask branch).[49] The FPN is a technique used to improve object detection and image segmentation by detecting and extracting features at different scales within the image.[50] In the FPN, CNNs are used to capture abstract and semantic features at deeper layers, while capturing detailed and spatial features at shallower layers in a bottom-up pathway. A top-down pathway up-samples the features from deeper layers to match the dimensions of the shallower layers, and lateral connections add the matching feature maps from the pathways allowing the flow of information between the levels of the hierarchy. Fig. 3 (left) shows the described FPN architecture, and Fig. 3 (right) shows the mask R-CNN head architecture using the ResNet backbone and the FPN extractor.

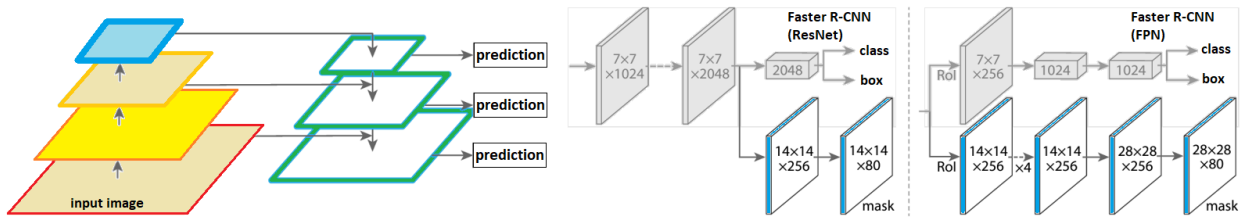

**Fig. 3** FPN architecture (left),[50] Mask R-CNN head architecture (right).[49] FPN = Feature pyramid network, R-CNN = Region-based convolutional neural network.

*2.3.3 SAM-based models*

SAM is a foundation model for promptable image segmentation released by FAIR in 2023. The three main components of SAM are illustrated in Fig. 4.[33] The image encoder uses powerful and scalable pre-trained masked autoencoder (MAE)[51] Vision Transformers (ViT).[52] In SAM the image encoder processes high-resolution inputs, it runs once per image and independently of the prompt encoder. The prompt encoder processes points, boxes, and text (sparse prompts) and masks (dense prompts). The sparse prompts are represented as positional encodings and the dense prompts are embedded using convolutions and are summed with the image embeddings. Finally, an efficient mask decoder maps the image and prompts embeddings using a bi-directional prompt self-attention and cross-attention embeddings updater. At the output, a dynamic linear classifier computes three mask probability foregrounds at each image location. The confidence score for each mask is also provided to avoid multiple valid image masks for ambiguous input prompts.[35]

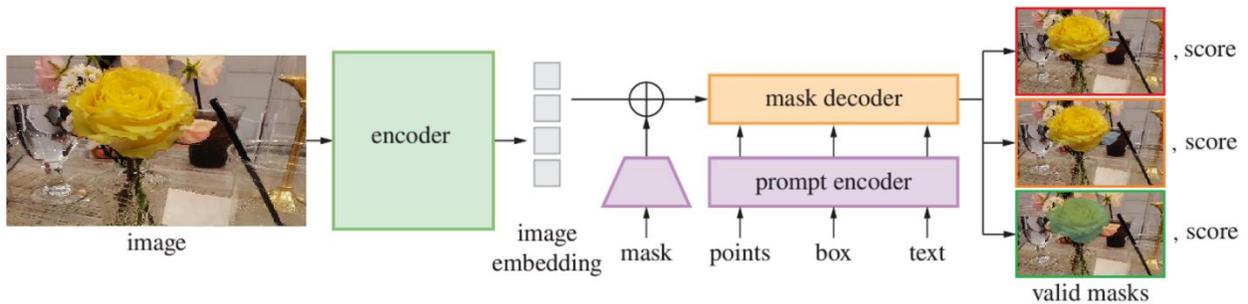

**Fig. 4** SAM architecture representation.[33]

SAM has provided excellent performance for segmenting general images such as landscapes, vegetation, images with humans and animals, and medical images including organs with clear boundaries. However, the performance of the model is significantly reduced when working with medical images that are noisy, have low contrast, and contain organs with weak boundaries.[35] A refined SAM foundation model called MedSAM improves the segmentation performance of the SAM model in medical images. MedSAM is a fine-tuned SAM on a large medical-labeled dataset with more than one million image pairs.

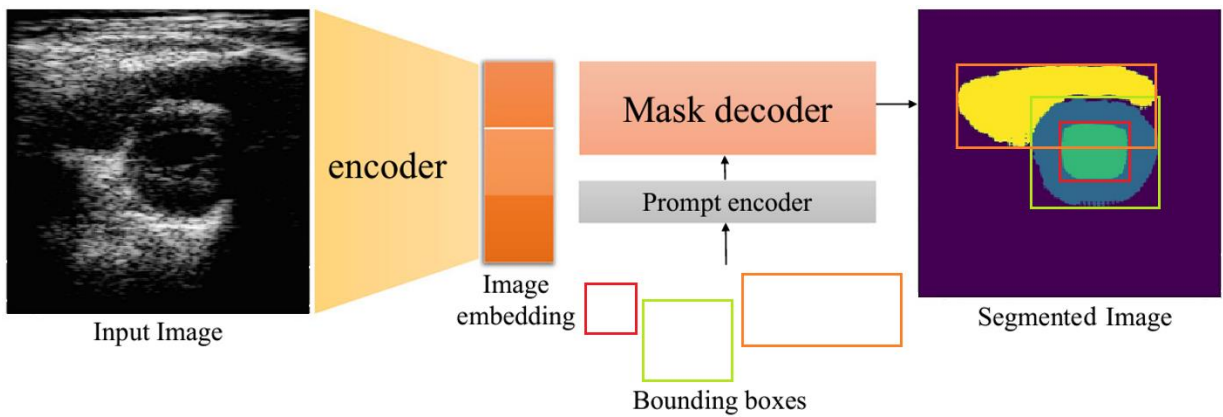

**Fig. 5** MedSAM architecture.[35]

As shown in Fig. 5, MedSAM resembles the SAM architecture using an image encoder and mask decoder, but the inputs for the prompt encoder were reduced to bounding boxes that were found to be more efficient in medical images that often require muti-organ or multi-object segmentation. Following the same procedure performed by the MedSAM developers, the MedSAM model was finetuned using our dataset of PSAX images, named US-MedSAM, and used in this study as the second variant of the original segment anything model developed by FAIR.

*2.4 Image Preprocessing*

The different architectures of the DL models used in this study require a specific format for the input images and labels. The dataset was initially preprocessed to meet the characteristics of the Unet-ResNet models and then adapted to the requirements of the Detectron2 and SAM-based models.

In the Unet-ResNet models, our PSAX-image dataset was examined for image consistency and minimum quality. In the review process, the exclusion criteria considered were (1) scans acquired with a transducer frequency different than 3.5 MHz, (2) scans acquired at a depth different than 16 cm, (3) images containing rib shadows covering more than ~30% of the image, (4) scans where there is partial or only one heart structure, and (5) scans for which the RF data was not saved during acquisition. The review process was performed by the clinical cardiologist in our team and yielded a final dataset of 1736 image-label pairs. The three manually delineated contours of the heart structures shown on Fig. 1 were flood-filled, converted to cartesian geometry, and used as image masks for training the DL models as depicted in Fig. 6. On each mask, the heart structures were labeled as left-ventricle (LV, purple color) for the flood-filled left-endocardial contour area, left-myocardium (LM, green color) for the flood-filled area between the left-endocardial and left-epicardial contours, and right-ventricle (RV, red color) for the flood-filled right-epicardium area. All image data and labels were resized to 256x256, the image data was saved as NumPy arrays, and the image masks as .bmp files.

The training data for Detectron2 must be in COCO format. The common objects in context – COCO format is a standard for object detection, instance segmentation, and keypoint detection. COCO annotations are usually in a JSON file with the file paths and dimensions of the images, including the bounding boxes, segmentation masks, and category labels. Annotations for Detectron2 were obtained by processing the image masks used in the Unet-ResNet models. The process involves creating an ID for each image, separately detecting the contour of each heart structure, drawing their bounding boxes, and saving the information in a JSON file. Examples of COCO annotations are shown in Fig. 7**,** the first two annotations include the bounding boxes and labels created for each heart structure. The Detectron2 visualization function randomized the color for each label and overlayed them over the input image.

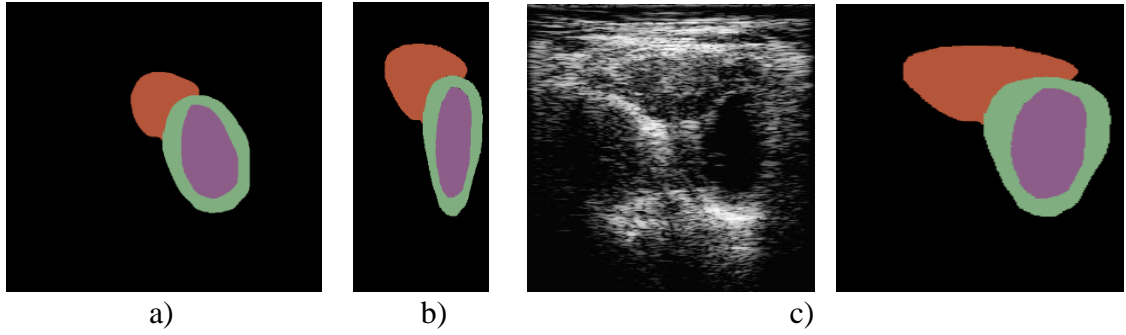

**Fig. 6** Dataset preprocessing for Unet-ResNet models. a) Flood-filled contours of End-Diastole frame, b) Conversion of the mask to cartesian, c) Image data and labels are resized to 256x256 pixels.

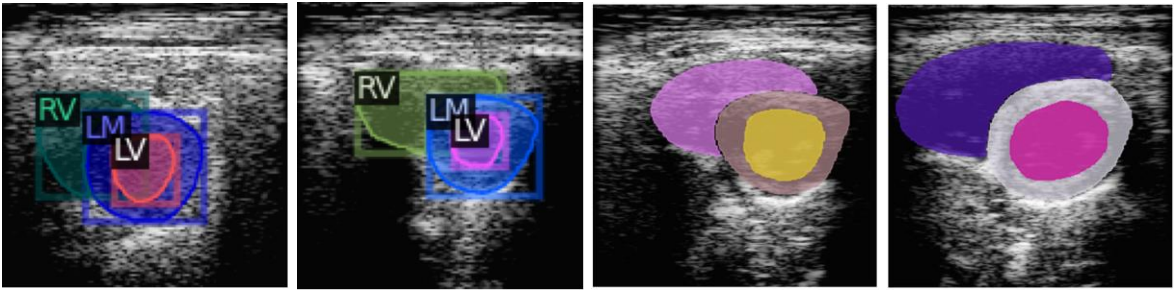

**Fig. 7** Examples of COCO annotations for the Detectron2 model. LV = left ventricle, LM = left myocardium, RV = right ventricle.

The preprocessing of the images for the SAM-based models required the extraction of the pixels of the cardiac structures from the Unet-ResNet model's image masks, the replacement of the color pixels with labels from 1 to 3, the resizing of the input images and masks to 1024x1024 pixels, and the *min-max* normalization of the input images. Input images and mask images were saved as NumPy arrays to be used in the training process.

*2.5 Validation Metrics*

The performance of the deep learning model for ventricular segmentation was evaluated using the Dice similarity coefficient (DSC), the difference in cross-sectional area (DCSA), and the Hausdorff distance (HD). DSC or Dice score measures the overlap between the area inside two contours, for example contour *G* (Gold standard) and contour *P (Prediction)*,[53] as shown in equation (1). When DSC is 1, contour *G* and contour *P* perfectly overlap. DCSA is the absolute difference between areas of the *G* and *P* contours. When the difference in area is 0, contour *G* and contour *P* are in good agreement.

$$DSC(G,P) = \frac{2(Area_G \cap Area_P)}{Area_G \cup Area_P} \quad (1)$$

HD measures the longest Euclidean distance between two points on the contour *G* and the contour *P*.[54]

$$HD(G,P) = max\big(h(G,P), h(P,G)\big) \quad (2)$$

where

$$h(G, P) = \max_{g \in G} (\min_{p \in P}(d(g,p))) \qquad (3)$$

$$h(P, G) = \max_{p \in P} (\min_{g \in G}(d(p,g))) \qquad (4)$$

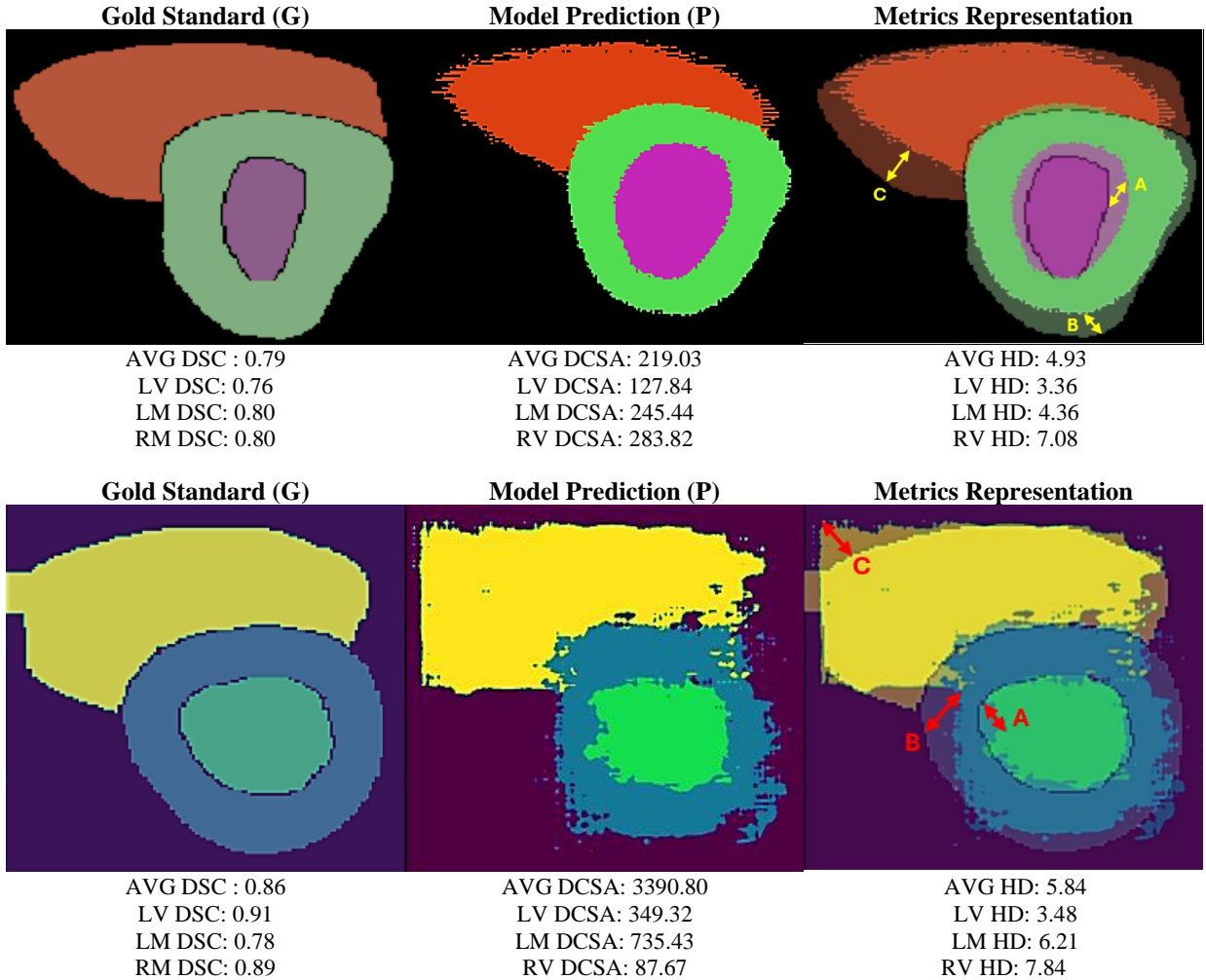

**Fig. 8** Comparison of the gold standard (G) and the segmentation predictions (P). The yellow and red arrows represent Euclidean distances between the G and P contours for LV (A), LM (B), and RV (C). AVG = Average, DSC = Dice similarity coefficient, DCSA = Difference in cross-sectional area, HD = Hausdorff distance, LV = left ventricle, LM = left myocardium, RV = right ventricle.

The function h(*G*, *P*) measures the distance between each point on *G* to the nearest point on *P*, then ranks points with the largest distance as the most mismatched points of *G*. The operation is repeated from points *P* to *G*. The final HD(*G*, *P*) is the maximum between h(*G*, *P*) and h(*P*, *G*) providing a measure of mismatch between the sets. Fig. 8 presents the gold standard used for training, the model prediction, and the overlap of the images to represent how metrics compare the gold standard and the model prediction.

## 3 Experiments

### 3.1 Setup

The dataset was randomly split into three subsets: a training set (70%), a validation set (10%), and a test set (20%). In addition, the training set was augmented with affine transformation techniques emulating the rocking movement of the ultrasound transducer as defined by the American Institute of Ultrasound in Medicine.[55] The rocking includes image rotations of -40°, -30°, -20°, -10°, 10°, 20°, 30° and 40° degrees. The training and validation sets were used to train the Unet-ResNet52, Unet-ResNet101, and Detectron2 models.

According to the architecture and training algorithm, each DL model performs additional image processing to the data set. The Unet-ResNet models normalize the image data using *z*-score normalization to reduce data range and enhance the accuracy of the trained model.[56] The normalization equation for each image (*I*) with $\mu$ as the global mean and $\sigma$ as the standard deviation is shown:

$$I_{z-score} = \frac{I - \mu}{\sigma} \quad (5)$$

The Detectron2 model required the registration of instances of the training and validation sets. The registration is performed using a Detectron2 library's function in which the JSON file with the annotations, and the input image paths are specified. Input images for the SAM-based models were already normalized using *min-max* normalization[57] during the preprocessing. In this normalization, the minimum value of the image (*I*) is subtracted from the image, and then divided by the range of the values in the image as shown:

$$I_{min-max} = \frac{I - min(I)}{max(I) - min(I)} \quad (6)$$

As described in section 2.2.1, SAM-based models are promptable DL models in which a point, a box, or text must be given as input to define the segmentation target. To handle this aspect, before the training, bounding boxes from the annotations are created for each label with a random displacement between 0 to 20 pixels of its original location over the x and y -axes. The fine-tuning process for the SAM-based models requires only training and test sets, for these models the training and validation sets were combined and used for training. The test set was used to evaluate the performances of the six models.

The Unet-ResNet models were implemented using TensorFlow, while the Detectron2 and SAM-based models were developed using PyTorch. The training of the models was optimized using the hyperparameters in Table 4 to mitigate overfitting. The standard categorical cross-entropy loss function for the SoftMax activation function was used as the cost function for multi-class classification.[58]

Table 4 Hyperparameters for the DL models training.

| Hyperparameter | Unet-Resnet | Detectron2 | SAM-based |
| --- | --- | --- | --- |
| Training/Validation/Testing Split (%) | 70/10/20 | 70/10/20 | 80/20 |
| Batch Size | 4 | 4 | 4 |
| Initial Learning Rate | 0.00008 | 0.00025 | 0.0001 |

| Learning Rate Drop | 0.25 | 0.25 | 0.01 |
| Early Stop | 20 | 20 | 20 |
| Number of Epochs | 100 | 1000 | 1000 |
| Patience | 2 | 2 | 2 |

## 3.2 Post-processing

After the prediction of the DL models is performed for the images in the test set, the generated probability maps must be post-processed to produce a final output. In the Unet-ResNet models, the probability maps for each cardiac structure are one-hot encoded and compared with the one-hot encoded mask by computing the DSC metric. The previous process also returns the areas of each object to compute the DCSA. Finally, the contours of the predictions and masks are extracted and used to calculate the HD metric.

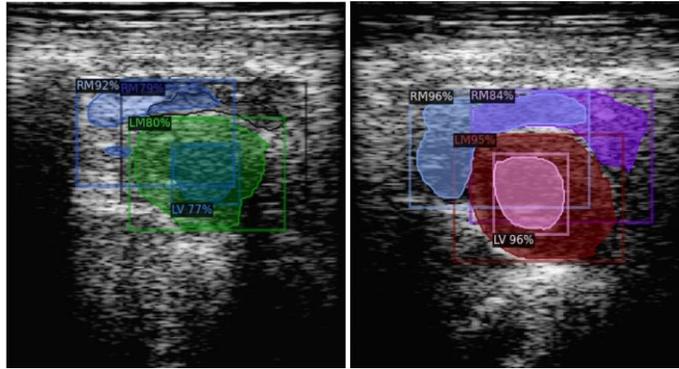

**Fig. 9** Multiple Detectron2 instance predictions of cardiac structures. LV = left ventricle, LE =left epicardium, RV = right ventricle.

Predictions performed by the Detectron2 models include the confidence computed for the model for each label as shown in Fig. 9. This is because the model was designed to perform instance segmentation and can identify more than one object of the same class in an image. That concept doesn't apply to the dataset of this study where each PSAX image contains one instance of each object. The Detectron2 predictions with the highest confidence were used as the prediction for each label and used to compute the DSC, HD, and DCSA metrics for each cardiac structure.

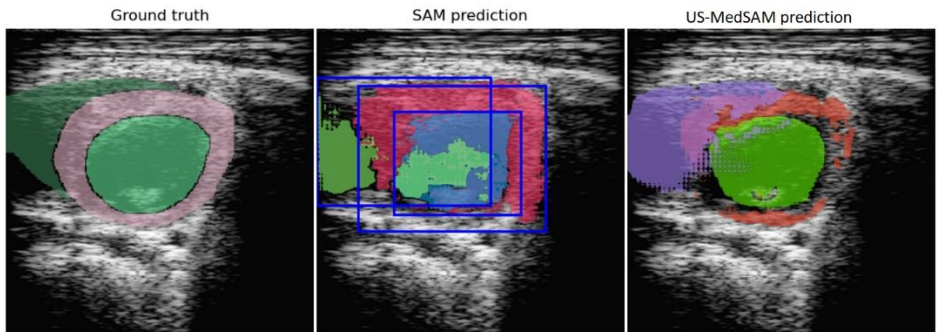

**Fig. 10** SAM-based models' predictions of cardiac structures without postprocessing.

As presented in Fig. 10, predictions from SAM-based models required handling the overlap between cardiac structures. The postprocessing performed for these outputs involved the extraction of pixels for each label and the application of image operations such as additions, subtractions, AND, XOR, and thresholding.

## 3.3 Results

This study's six DL models were evaluated using a test set of 349 images. The Unet-ResNet models were demonstrated to be robust by generating predictions containing the 3 expected cardiac structures for the entire test set, while Detectron2 did not perform as well. Examples are presented in Fig. 11 for some images of the test set where the Detectron2 and the MedSAM models did not predict some cardiac structures.

| Test Image # | 126 | 345 | 51 |
|---|---|---|---|
| Input Image | 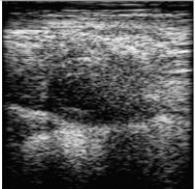 | 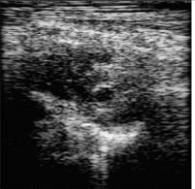 | 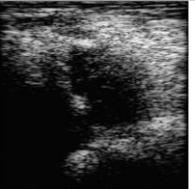 |
| Ground truth | 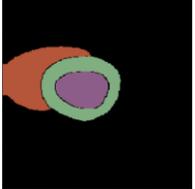 | 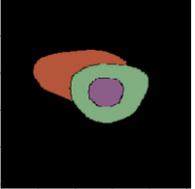 | 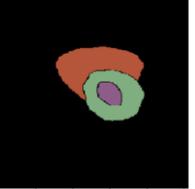 |
| UnetResnet50 | 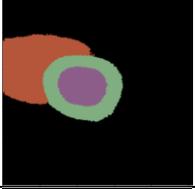 | 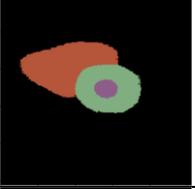 | 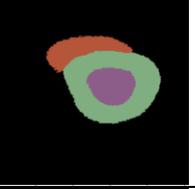 |
| UnetResnet101 | 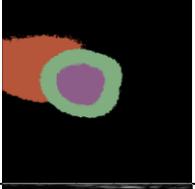 | 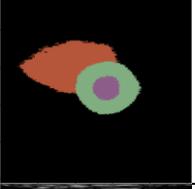 | 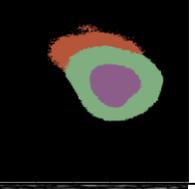 |
| Detectron2 | 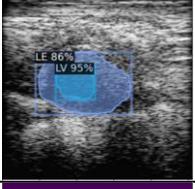 | 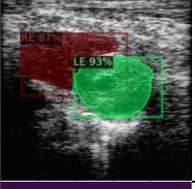 | 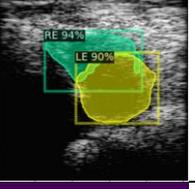 |
| SAM | 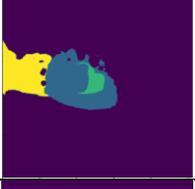 | 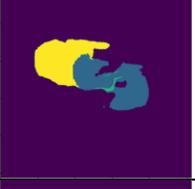 | 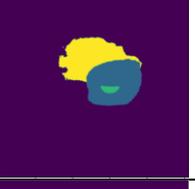 |
| MedSAM | 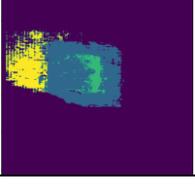 | 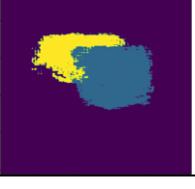 | 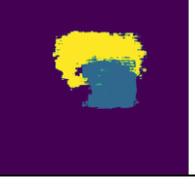 |

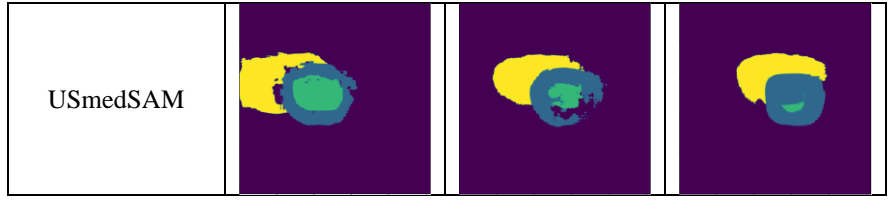

USmedSAM

**Fig. 11** Examples of incomplete predictions made by the deep learning models. LV = left ventricle, LM = left myocardium, RV = right ventricle.

**Table 5** Statistics of validation metrics for DL models. DSC = dice similarity coefficient, DCSA = difference in cross-sectional area, HD = Hausdorff distance, LV = left ventricle, LM = left myocardium, RV = right ventricle.

| Metric | DSC | | | DCSA | | | HD | | |
|---|---|---|---|---|---|---|---|---|---|
| Object | LV | LM | RV | LV | LM | RV | LV | LM | RV |
| **UnetResnet50** | **0.77** ± 0.11 | 0.85 ± 0.09 | 0.82 ± 0.11 | **122.65** ± 112.27 | **70.83** ± 66.47 | **150.27** ± 144.30 | 4.35 ± 2.19 | 3.54 ± 1.92 | 7.28 ± 4.19 |
| **UnetResnet101** | **0.78** ± 0.10 | **0.86** ± 0.08 | **0.83** ± 0.10 | **114.07** ± 96.87 | **63.17** ± 57.57 | **140.71** ±125.02 | **4.20** ± 2.04 | **3.38** ± 1.69 | **7.20** ± 4.33 |
| **Detectron2** | 0.73 ± 0.10 | 0.84 ± 0.09 | 0.77 ± 0.15 | 127.83 ± 92.86 | 71.79 ± 63.15 | 150.58 ± 135.37 | **1.75** ± 0.74 | **1.35** ± 0.70 | **3.23** ± 1.86 |
| **SAM** | 0.50 ± 0.14 | 0.53 ± 0.23 | 0.77 ± 0.11 | 2009.74 ± 912.74 | 3301.05 ± 953.68 | 2323.62 ± 915.04 | 11.85 ± 2.93 | 11.00 ± 5.81 | 10.38 ± 5.17 |
| **medSAM** | 0.65 ± 0.12 | 0.67 ± 0.22 | 0.72 ± 0.14 | 2128.03 ± 973.28 | 1821.63 ± 989.33 | 2942.48 ± 918.90 | 12.04 ± 4.03 | 6.95 ± 3.64 | 14.89 ± 6.43 |
| **USmedSAM** | 0.75 ± 0.09 | **0.89** ± 0.08 | **0.88** ± 0.03 | 1971.55 ± 926.35 | 649.89 ± 761.32 | 1135.25 ± 962.74 | 8.20 ± 3.57 | 4.04 ± 4.30 | 7.74 ± 4.36 |

To compare the performance of the six DL models, 80 input images for which not all the models predicted the 3 cardiac structures were excluded from the analysis. The average and standard deviation for each metric were computed and the results are shown in Table 5. The Unet-ResNet models provided better performance for all metrics (values in bold), where the Unet-ResNet101 got the best scores for DSCA and LV-DSC and the second-best scores (values in blue) for the other metrics and cardiac structures. The performance of the Detectron2 model was superior in terms of the HD. This means that it predicted contours with a smaller pixel difference than the ground truth, but the performance for the other metrics was below the Unet-ResNet models. The US-MedSAM provided the higher DSC performance for the LM and RV cardiac structures indicating its predictions were in the most correct spatial position, but this model provided high values in the DSCA due to the lack of pixel predictions as shown in Fig. 11 and Fig. 12.

Table 6 presents the test image numbers, and best DSC and HD scores computed for each DL model. In the table we can see how all the models provided DSC metrics above 0.90 for some cardiac structures, and HD values less than 5 pixels. The result is qualitatively presented in Fig. 12 including the input image, ground truth, and the best prediction for each model according to the higher DSC metric value.

**Table 6** Best DSC and HD scores for each DL model. DSC = dice similarity coefficient, HD = Hausdorff distance, Img. = image, LV = left ventricle, LM = left myocardium, RV = right ventricle, AVG = average.

| Metric | DSC | | | | | HD | | | | |
|---|---|---|---|---|---|---|---|---|---|---|
| Object | Test Img. | LV | LM | RV | AVG | Test Img. | LV | LM | RV | AVG |
| **UnetResnet50** | 308 | 0.94 | 0.98 | 0.97 | 0.96 | 308 | 0.95 | **0.75** | **1.38** | **1.03** |

| | | | | | | | | | |
|---|---|---|---|---|---|---|---|---|---|
| **UnetResnet101** | <u>311</u> | **0.94** | **0.97** | **0.94** | **0.95** | <u>311</u> | **0.75** | 0.95 | 2.68 | 1.46 |
| **Detectron2** | <u>327</u> | 0.87 | 0.95 | 0.92 | 0.91 | <u>227</u> | **0.66** | **0.45** | **1.22** | **0.78** |
| **SAM** | <u>73</u> | 0.77 | 0.90 | 0.87 | 0.85 | <u>266</u> | 4.22 | 2.55 | 4.41 | 3.73 |
| **medSAM** | <u>311</u> | 0.78 | 0.91 | 0.89 | 0.86 | <u>341</u> | 5.85 | 2.72 | 5.63 | 4.73 |
| **USmedSAM** | <u>136</u> | 0.92 | 0.94 | 0.93 | 0.93 | <u>212</u> | 2.77 | 1.86 | 2.75 | 2.46 |

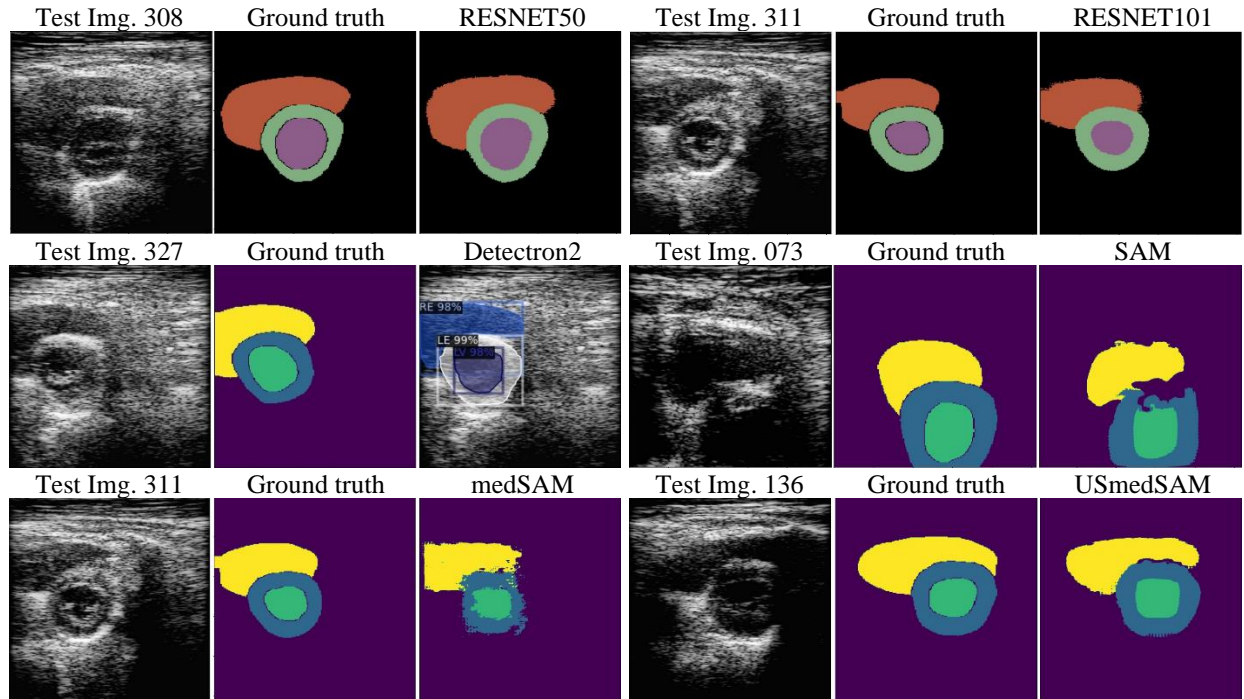

**Fig. 12** Predictions with best DSC scores for each DL model. LV = left ventricle, LM = left myocardium, RV = right ventricle.

## 4 Discussion

The main objective of this study was to evaluate SOTA DL models in the segmentation of the LV and the RV in 2D short-axis echocardiographic images. Deep learning models, including the Unet-ResNet50, Unet-ResNet101, Detectron2, SAM, MedSAM, and US-MedSAM were implemented, trained, fine-tuned, and used for performance comparisons.

Among the three target objects, the right epicardium (RE) contour or whole RV area is generally underrepresented compared to the left ventricle in cardiac image segmentation, especially in echocardiography. RV segmentation suffers from many limiting factors, including irregular and inconsistent geometry of the RV, similarity in grayscale values of the chamber and myocardial border of the RV, and poor quality of the RV appearance in the ultrasound images. As a result, the RV delineation by experts even in a manual manner produces variability. The 2 cases in Fig. 13 illustrate differences in the determination of the shape and length of the RV contour among three experts. Having a higher average DSCA and higher Hausdorff distance of the RV compared to the other objects also affirms the difficulty in segmentation of the RV. Compared to the DSC of the segmentation results of the cardiac structures, the DSCs of the LV (left endocardium) predictions are slightly lower. This is despite the qualitative results shown in Fig.

11 presenting a good agreement. In the Unet-ResNet models, the area differences resulting from test images #345 and #51 impact this metric.

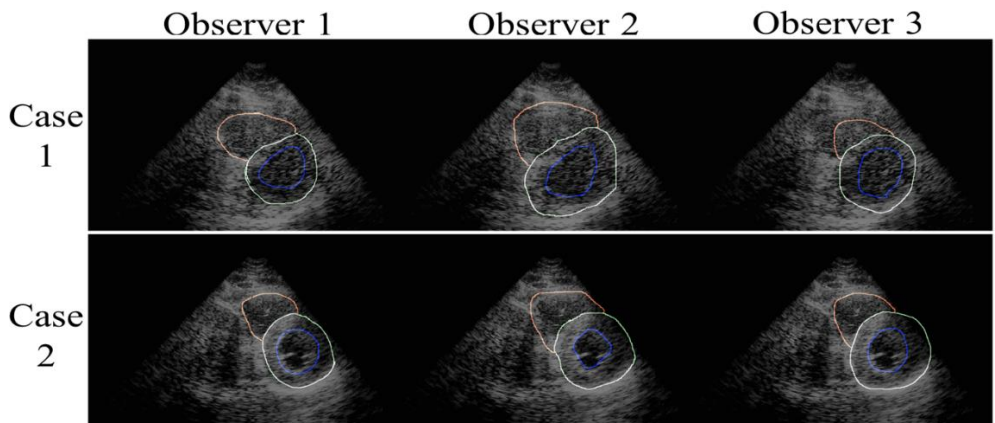

**Fig. 13** Differences in drawing contours in 2 echocardiogram cases from experts.

The SAM model was able to make predictions in our dataset without any fine-tuning. The incomplete predictions shown in Fig. 11 and best predictions in Fig. 12 demonstrated that, with some improvements, the model performance could be increased. This was also shown to be the case when using the fine-tuned models MedSAM and US-MedSAM. The original Detectron2 model didn't provide predictions for test images before the fine-tuning. Its performance was close to the fine-tuned US-MedSAM for the DSC metric and above the fine-tuned US-MedSAM for the DSCA and HD metrics. As expected, the models with the best performance (Unet-ResNets) achieved consistent metric values for the test images, while the irregularities in the predictions of the other models yielded the best metric scores in different images. Some examples are presented in Table 5. The highest DSC score for Detectron2 was provided for test image #327 but the highest HD was in test image #227. Another example is from US-MedSAM with a higher DSC for test image #136 and higher HD for test image #212. In terms of the cardiac structures' areas, the models predominantly tended to underestimate the LV, as shown in Fig. 11. This tendency was more pronounced in the SAM-based models, where predictions for this structure were often absent. Consequently, the absence of LV predictions led to an overestimation of the LM, resulting in higher DCSA values in both the original SAM and MedSAM models.

This study still has some notable limitations. One challenge in applying deep learning to medical image segmentation is the requirement of a relatively large, annotated training dataset. As presented in Table 1, SOTA models for image segmentation are trained with datasets sometimes including ~1 million images. In this study, taking advantage of the transfer learning and fine-tuning techniques, the previous knowledge of the initial weights for Detectron2 and SAM models was used to retrain the models with our PSAX-echo dataset. Models were trained and fine-tuned without increasing the image dataset with public datasets such as CAMUS,[42] or EchoNet-Dynamic[43] because they mainly contain parasternal long-axes views (PLAX) and don't provide the additional RF data required for future research in CAT detection and quantification using echocardiograms.

# 5 Conclusion

In this paper, an evaluation of specific-domain and general-domain SOTA segmentation DL models was presented. The architectures of the Unet-ResNet50, Unet-ResNet101, Detectron2, SAM, MedSAM, and the US-MedSAM models were applied to ventricular segmentation in short-axis echocardiograms. This study successfully segmented the right ventricle (RV) from a private parasternal short-axis (PSAX) echocardiogram dataset, a perspective that has not been extensively explored in previous research. Our results demonstrated comparable performance to models trained on public datasets with standard echocardiographic views, such as A2C, A3C, and A4C.

The DSC scores and HD values obtained for the three target structures highlight the potential of our deep learning (DL) framework for automated segmentation of cardiac structures, particularly the RV, which has been less frequently studied in prior work. Additionally, this study underscores the superior performance of domain-specific DL models compared to general-domain models, even when fine-tuning is applied to the latter.

Future research could focus on comparing the performance of our framework with models trained on public datasets such as CAMUS and EchoNet. Enhancing our PSAX dataset with advanced data augmentation techniques and acquiring echocardiographic and radiofrequency (RF) data from all standard imaging views would enable a more comprehensive approach to segmentation, encompassing all visible cardiac structures. Moreover, evaluating model performance on full cardiac cycles could provide valuable insights into the diagnosis of cardiac cycle dysfunctions.

**Disclosures**

All authors declare to have no conflicts of interest.

**Code, Data, and Materials Availability**

The data utilized in this study were obtained from the Biomedical Imaging Research Lab – BIRL at SIUE (https://siue-biomedicalimaginglab.com/). Data are available from the authors upon request, and with permission from BIRL leader Dr. Jon Klingensmith. The code and sample data are publicly available at: https://github.com/SIUE-BiomedicalImagingResearchLab/SOTA-DL-EchoSegmentation

**Acknowledgments**

This work was supported by the American Heart Association – AHA (Predoctoral fellowship https://doi.org/10.58275/AHA.24PRE1188140.pc.gr.190654 to Julian R. Cuellar), and partially by the National Heart, Lung, and Blood Institute of the National Institutes of Health under Award Number R15HL145576.

2256. Reinhold JC, Dewey BE, Carass A, Prince JL. Evaluating the Impact of Intensity Normalization on MR Image Synthesis. Published online December 11, 2018. doi:10.48550/arXiv.1812.04652
57. Cuellar JR, Karlapalem A, Umbaugh SE, Marino DJ, Sackman J. Detection of syrinx in thermographic images of canines with Chiari malformation using MATLAB CVIP toolbox GUI. In: *Proc. SPIE 11004, Thermosense: Thermal Infrared Applications XLI, 1100405*. ; 2019. https://doi.org/10.1117/12.2519112
58. Hung WC, Tsai YH, Liou YT, Lin YY, Yang MH. Adversarial Learning for Semi-Supervised Semantic Segmentation. Published online July 24, 2018. Accessed August 5, 2023. http://arxiv.org/abs/1802.07934
22

**Julian Cuellar** is a Ph.D. in engineering sciences candidate in the cooperative Ph.D. program at Southern Illinois University Carbondale and Southern Illinois University Edwardsville. He received his M.Sc. degree in electrical engineering from Southern Illinois University Edwardsville in 2019.

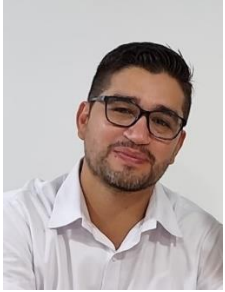

**Vu Dinh** is a Sr. Research Specialist at the Pathak lab division of Cancer Imaging Research, department of radiology, Johns Hopkins Medicine, Baltimore, Maryland. He got his M.Sc. in electrical engineering in the Department of Electrical and Computer Engineering at Southern Illinois University Edwardsville in 2023, where he also received his bachelor's degree in electrical engineering in 2021.

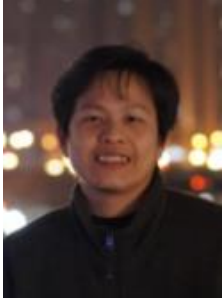

**Manjula Burri** is a cardiologist affiliated with the Columbus Regional Hospital in Columbus, IN, USA. She has over 20 years of experience. Her specialties include Cardiovascular Disease and Internal Medicine, and she is Board certified in Cardiovascular disease. She received her medical residency from Mt. Sinani School of Medicine at Queens Hospital Center, Jamaica, NY, USA.

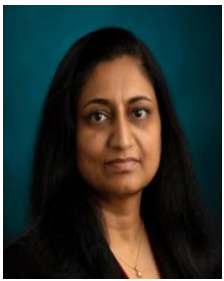

**Julie Roelandts** works for the Diagnostic Medical Sonography department at St. Louis Community College. She received her associate's degree in Diagnostic Medical Sonography from Cuyahoga Community College, Cleveland, Ohio in 2003.

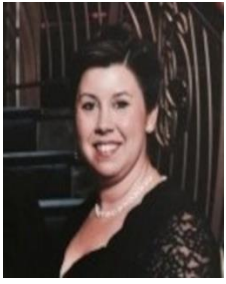

**James Wendling** is a program director at St. Louis Community College in the diagnostic medical sonography program. He received his bachelor's degree in exercise science from Western Illinois University, Macomb, Illinois in 1995.

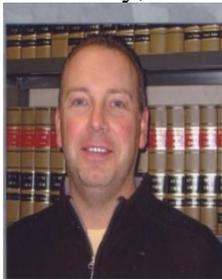

**Jon D. Klingensmith** is an Assoc. Professor in the Department of Electrical and Computer Engineering at Southern Illinois University Edwardsville. He received his PhD in Biomedical Engineering from Case Western Reserve University.

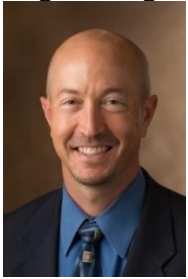

**List of Figures**

**Fig. 1** Gold standard development. a) Echocardiogram with 80 temporal frames, b) Identification of End-Diastole frame, c) Identification of End-Systole frame. Experts traced the left myocardium, left epicardium, and right epicardium using blue, green, and red colors respectively.
**Fig. 2** U-Net architecture (left), ResNet residual block (center), and decoder block (right).
**Fig. 3** FPN architecture (left),[50] Mask R-CNN head architecture (right).[49] FPN = Feature pyramid network, R-CNN = Region-based .
**Fig. 4** SAM architecture representation.[33]
**Fig. 5** MedSAM architecture.[35]
**Fig. 6** Dataset preprocessing for Unet-ResNet models. a) Flood-filled contours of End-Diastole frame, b) Conversion of the mask to cartesian, c) Image data and labels are resized to 256x256 pixels.
**Fig. 7** Examples of COCO annotations for the Detectron2 model.

**Fig. 8** Comparison of the gold standard (G) and the segmentation predictions (P). The yellow and red arrows represent Euclidean distances between the G and P contours for LV (A), LM (B), and RV (C). AVG = Average, DSC = Dice similarity coefficient, DCSA = Difference in cross-sectional area, HD = Hausdorff distance, LV = left ventricle, LM = left myocardium, RV =
**Fig. 9** Multiple Detectron2 instance predictions of cardiac structures.
**Fig. 10** SAM-based models' predictions of cardiac structures without postprocessing.
**Fig. 11** Examples of incomplete predictions made by the deep learning .
**Fig. 12** Predictions with best DSC scores for each DL model.
**Fig. 13** Differences in drawing contours in 2 echocardiogram cases from experts.

**List of Tables**

**Table 1** Characteristics of SOTA image segmentation models.
**Table 2** Parasternal Short Axis image acquisition details.
**Table 3** Layer details of the Unet-ResNet50, and the Unet-ResNet101 architectures. Details of ResNet residual and decoder blocks are depicted in Fig. 2 (center) and Fig. 2 (right).
**Table 4** Hyperparameters for the DL models training.
**Table 5** Statistics of validation metrics for DL models.
**Table 6** Best DSC and HD scores for each DL model.